\documentclass[twocolumn,showpacs,preprintnumbers,amsmath,amssymb, prb, reprint]{revtex4-1}

\usepackage{stmaryrd}
\usepackage{txfonts}
\usepackage{amssymb}
\usepackage{mathrsfs}
\usepackage{graphicx}
\usepackage{dcolumn}
\usepackage{bm}
\usepackage{epsfig}
\usepackage{color}                    
\usepackage{hyperref}                 

\newcommand{\beq}{\begin{equation}}
\newcommand{\eeq}{\end{equation}}
\newcommand{\bes}{\begin{subequations}}
\newcommand{\ees}{\end{subequations}}
\newcommand{\bea}{\begin{eqnarray}}
\newcommand{\eea}{\end{eqnarray}}
\newcommand{\ba}{\begin{array}}
\newcommand{\ea}{\end{array}}
\newcommand{\beqn}{\begin{eqnarray*}}
\newcommand{\eeqn}{\end{eqnarray*}}

\newcommand{\f}[2]{\frac{#1}{#2}}

\newcommand{\om}{\omega}

\newcommand{\la}{\langle}
\newcommand{\ra}{\rangle}

\def\nn{\nonumber}

\newlength{\sizeonefig}
\newlength{\sizetwofig}
\begin{document}
\preprint{\href{http://dx.doi.org/10.1088/0953-8984/25/32/325701}{S.-Z. Lin and D. Roy, J. Phys.: Condens. Matter {\bf 25}, 325701 (2013).}}

\title{Role of kinetic inductance in transport properties of shunted superconducting nanowires}
\author{Shi-Zeng Lin$^1$ and Dibyendu Roy$^2$}
\affiliation{$^1$Theoretical Division, Los Alamos National Laboratory, Los Alamos, New Mexico 87545, USA} 
\affiliation{$^2$Theoretical Division and Center for Nonlinear Studies, Los Alamos National Laboratory, Los Alamos, New Mexico 87545, USA}

\begin{abstract}
Recently transport measurements have been carried out in resistively shunted long superconducting nanowires [M. W. Brenner {\it et. al.}, Phys. Rev. B {\bf{85}}, 224507 (2012)]. The measured voltage-current ($V$-$I$) characteristics was explained by the appearance of the phase slip centers in the shunted wire, and the whole wire was modeled as a single Josephson junction. The kinetic inductance of the long nanowires used in experiments is generally large. Here we argue that the shunted superconducting nanowire acts as a Josephson junction in series with an inductor. The inductance depends on the length and the cross section of the wire. The inclusion of inductance in our analysis modifies the $V$-$I$ curves, and increases the rate of switching from the superconducting state to the resistive state.  The quantitative differences can be quite large in some practical parameter sets, and might be important to properly understand the experimental results. Our proposed model can be verified experimentally by studying the shunted superconducting nanowires of different lengths and cross sections.  
\end{abstract}
\pacs{74.78.Na, 74.25.F-, 74.25.Sv, 74.40.-n}
\date{\today}
\maketitle

\section{Introduction}
 Superconducting nanowires (SCNWs) of radius of several nanometers and length of hundreds of nanometers are under an intense current research for studying various interesting physical phenomena and potential technological applications including the fluctuations in superconductors, the superconducting qubits and the current standards \cite{Bezryadin08}. SCNWs also have promising applications in the single-photon detection. \cite{Ilin2000,Goltsman2001,Natarajan2012}  The voltage-current ($V$-$I$) characteristics of the SCNWs are measured by several groups, and a hysteresis has been observed. Upon increasing the bias current, the SCNW switches to a resistive state at a critical current $I_{\rm sw}$. When one reduces the bias current from the resistive state, the nanowire remains in the resistive state until the current is below a retrapping current $I_{\rm r}$. The $V$-$I$ curves of a SCNW are very similar to those of an underdamped Josephson junction (JJ).

By numerically solving the generalized time-dependent Ginzburg-Landau equations, it was shown in Refs. \onlinecite{Vodolazov03, Michotte04} that the hysteresis in the SCNWs can be explained by allowing the occurrence of a phase slip center (PSC). The superconducting phase in a PSC evolves with time according to the ac Josephson relation, and the superconductivity is suppressed, while the phase at the two sides of the PSC remains constant \cite{Little67,Langer67,McCumber70}. The PSC provides a mechanism for survival of superconductivity even in the presence of an electric field inside the superconductor. The SCNW with a PSC  can be modeled as a JJ. When the SCNW is driven by a dc bias current, a hysteresis shows up in the observed $V$-$I$ curves. However for a voltage biased case,  the reported $V$-$I$ curves are S-shaped \cite{Vodolazov03}. For a review on the PSC, please check Ref. \onlinecite{Ivlev84}.

The transport measurements are generally performed with the free-standing nanowires \cite{Shah08, Pekker09, Sahu09, Brenner12}. The exchange of heat from the free-standing SCNWs is weak because the cooling occurs only at the two ends of the nanowire where they are attached to superconducting electrodes. The self-heating thus becomes important for these nanowires. The hysteretic $V$-$I$ curves were obtained in a theoretical analysis by Tinkham \emph{et. al.}\cite{Tinkham03} by using the measured temperature $T$ dependent linear resistance $R(T)$ as an input in solving the heat flow equation through the SCNWs. At some current interval (between $I_{\rm r}$ and $I_{\rm sw}$), there are two possible states with one being the heating dominated resistive state and the other superconductivity dominated state. A large fraction of the nanowire becomes normal in the heating dominated resistive state while only a small fraction of the wire is normal in the superconductivity dominated state. Therefore the nanowire shows the hysteretic $V$-$I$ curves. The self-heating model was further advanced in Ref. \onlinecite{Pekker09}.

The switching current $I_{\rm sw}$ becomes stochastic in the presence of thermal or/and quantum fluctuations. A distribution of $I_{\rm sw}$ is obtained in experiments by repeating the measurements of $I_{\rm sw}$. It has been demonstrated for the free-standing unshunted nanowires that the width of the distribution of $I_{\rm sw}$ increases with a decreasing  temperature \cite{Shah08, Pekker09, Sahu09}. This counter-intuitive result was argued from the self-heating model. The unshunted SCNW is abounded with thermal phase slips at relatively high temperatures. The phase slip fluctuations induce resistance which causes Joule heating in the bulk, and the temperature of SCNW increases above the electrode temperature. When the electrode temperature is reduced, the self-heating caused by the dissipative phase slips becomes even more important as a result of the reduced heat capacity and heat conductivity. It was also observed that the retrapping at $I_{\rm r}$ is not stochastic for the unshunted SCNW within the experimental resolution.

The $V$-$I$ characteristics of the nanowires with an external shunt resistor have been measured last year \cite{Brenner12}. The $V$-$I$ curves of a shunted SCNW changes dramatically in comparison to that of an unshunted SCNW. While the mean switching current depends weakly on the external shunt resistance, the mean retrapping current increases with a decreasing shunt resistance. The mean retrapping current merges with the mean switching current at a threshold shunt resistance, and the hysteresis in $V$-$I$ curves vanishes. The distribution width of the switching current reduces with a decreasing shunt resistance or temperature in the low temperature region. All these observations cannot be explained by the self-heating model, which is only valid for the unshunted nanowires. The self-heating is reduced for a shunted nanowire because when there is a phase slip in the nanowire, the resistance of the wire is finite which causes a redistribution of the applied bias current in the shunt circuit. The authors of Ref. \onlinecite{Brenner12}  have modeled the SCNW with a PSC as a JJ. The derived $V$-$I$ curves and the distributions of the switching current based on their model of a JJ with an external shunt show reasonable qualitative agreement with the experimental results.

The role of the rest part of the nanowire excluding the PSC at a weak link  was completely neglected in the RCSJ type modeling of Ref. \onlinecite{Brenner12}. After a PSC develops in the SCNW, the bias current is redistributed between the nanowire and the shunt resistor. Then an electric field is induced in the SCNW resulting from the time-dependent supercurrent. In this work, we show that the rest part of the SCNW apart from the PSC acts as a kinetic inductor due to the kinetic energy of Cooper pairs, which introduces a retardation in the dynamics of the superconducting phase. The value of the inductance depends on the length and the cross section of the wire. The kinetic inductance affects the retrapping current, and also transforms the distribution of the switching current. Furthermore, the kinetic inductance reduces the self-heating effect in the nanowire. Thus, we conclude that the kinetic inductance of the SCNW cannot be neglected. 

The remaining part of the paper is organized as follows. In Sec. II, we derive the quantum circuit equations for our model. In Sec. III, the $V$-$I$ curves without noise at zero temperature are calculated. In Sec. IV, the effect of the kinetic inductance on the rate of the switching from the superconducting state is investigated. A short discussion on our results and some comparisons with other related recent studies are presented in  Sec. V. 

\begin{figure}[t]
\includegraphics[width=\columnwidth]{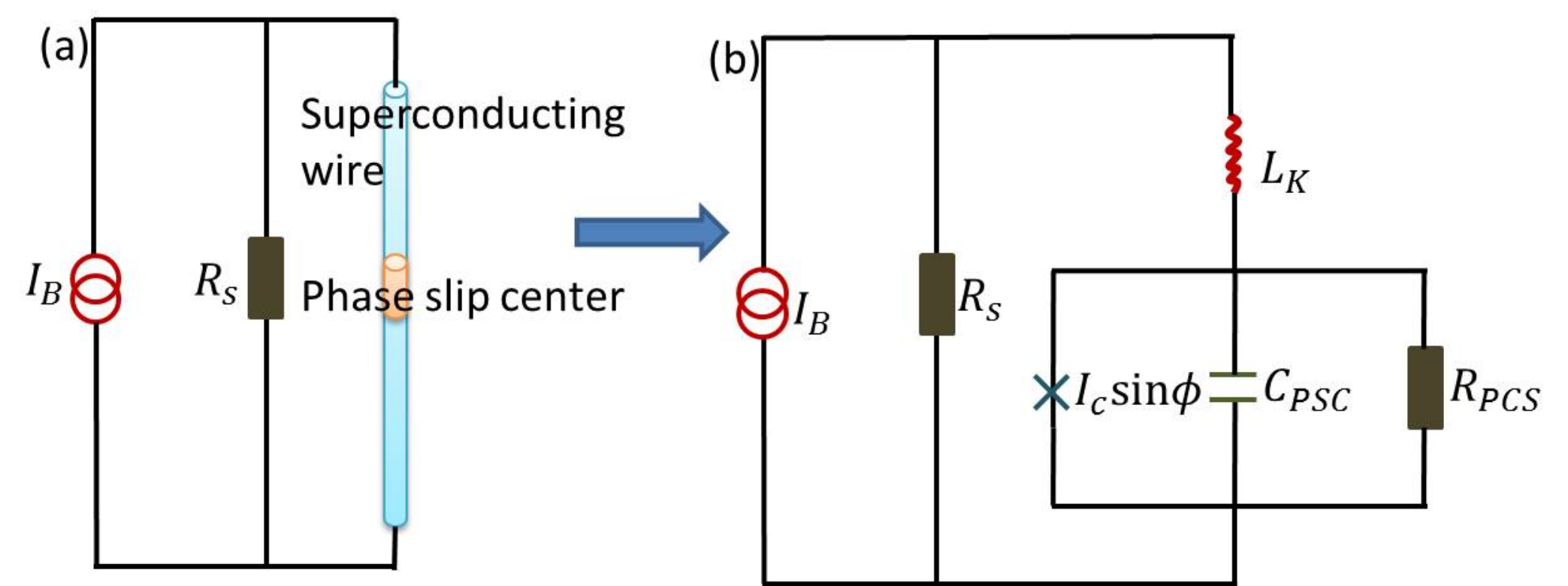}
\caption{(color online) (a) A schematic of a dc current biased superconducting nanowire with an external shunt resistor. The nanowire with an external shunt can be represented by an equivalent circuit shown in (b). The nanowire is modeled by a resistively capacitively shunted Josephson junction (RCSJ) in series with a kinetic inductor.}
\label{f1}
\end{figure}

\section{Quantum circuit and nonlocal damping}
We model the SCNW with a PSC as a JJ connected in series with an inductor, as shown schematically in Fig. \ref{f1}. The resistance of the JJ is due to the phase slips, and can be measured experimentally. The inductor describes the rest part of the superconducting region excluding the PSC. The inductance can be obtained from the London equation which states that the time-varying supercurrent density $\mathbf{J}_{\rm s}$ induces an electric field $\mathbf{E}$ inside the superconductor $\mathbf{E}=4\pi\lambda^2 \partial_t\mathbf{J}_{\rm s}/c^2$ where $\lambda$ the London penetration depth. Therefore, the kinetic inductance of the SCNW is $L_{\rm K}\equiv V/\partial_t I_{\rm s}=4\pi\lambda^2 l/(c^2 A)$, where $I_{\rm s}=J_s A$, $l$ is the length of the wire and $A$ is the cross section area of the nanowire. When the bias current is smaller than $I_{\rm sw}$, the whole bias current flows through the SCNW in the absence of phase slips. In the presence of phase slips in the nanowire, the resistance of the wire becomes finite and the current redistributes in the shunt circuit. The kinetic inductance $L_{\rm K}$ introduces a time scale for the redistribution  $\tau_1=L_{\rm K}/(R_{\rm s}+R_{\rm PSC})$ where $R_{\rm PSC}, R_{\rm s}$ are the resistance of the PSC and the external shunt resistor respectively. As the phase slip disappears, the nanowire recovers to the superconducting state. The current flows fully again into the nanowire, and the time scale for this redistribution is $\tau_2=L_{\rm K}/R_{\rm s}$. In the experiment \cite{Brenner12}, $R_{\rm PSC}> R_{\rm s}$, so $\tau_2> \tau_1$. Thus there is a sufficient spare time for the nanowire to cool down before the bias current redistributes into the wire. Therefore, the inductance reduces the heating effect in the SCNWs with a small external shunt resistor. The reduced self-heating effect due to the kinetic inductance of the SCNWs was demonstrated experimentally in NbN superconducting wires for single photon detection \cite{Kerman06, Yang07}. On the other hand, the geometric inductance of the loop $L_G=4\pi A_L/(c^2 l_L)$ is negligible in comparison with the kinetic one when $\lambda^2/A\gg 1$, where $A_L$ is the area and $l_L$ is the length of the loop.  

The SCNW shunted by an external resistor can be represented by an effective circuit as shown in Fig. \ref{f1}.  We use the resistively capacitively shunted Josephson junction (RCSJ) to model the PSC \cite{Tinkham96}. It has been shown in Ref. \onlinecite{Cawthorne98} that there are kinks in the $V$-$I$ curves of a resistively shunted JJ in the presence of an inductance in series with the external shunt resistance, and such kinks have been observed in the experiment with resistively shunted nanowire \cite{Brenner12}. We here study the role of kinetic inductance intrinsic to the nanowire, i.e., the inductance is in series with the PSC modeling the nanowire. Nevertheless the equations of motion for the superconducting phase and current in our case have the same form as those in Ref. \onlinecite{Cawthorne98}.

 In the following we provide a fully quantum mechanical description of the circuit in Fig. \ref{f1}(b) and derive the equations of motion for the superconducting phase and the current through the PSC \cite{Guichard10, Mooij06}. Here, $\hat{\phi}$ is a phase difference across the PSC, $\hat{Q}$ is the canonically conjugate charge at the PSC with the commutation relation, $[\hat{\phi},\hat{Q}]=2ie$ where $e$ is the charge of an electron. The phase differences across the  shunt resistance and the inductance are $\hat{\phi}_{\rm R}$ and $(\hat{\phi}_{\rm R}-\hat{\phi})$ respectively. The capacitance of the PSC is $C_{\rm PSC}$. The applied bias current is $I_{\rm B}$, and $\hat{I}_1$ is an operator for the current through the inductance (and the PSC). It is given by $\hat{I}_1=\hbar(\hat{\phi}_{\rm R}-\hat{\phi})/(2eL_{\rm K})$. The full Hamiltonian reads $\hat{H}=\hat{H}_{\rm 0}+\hat{H}_{\rm BP}+\hat{H}_{\rm BS}$, where  
\bea
H_{\rm 0}&=&\f{\hat{Q}^2}{2C_{\rm PSC}}-\f{\hbar I_{\rm c}}{2e}\cos \hat{\phi}+\Big(\f{\hbar}{2e}\Big)^2\f{(\hat{\phi}_{\rm R}-\hat{\phi})^2}{2L_{\rm K}}-\f{\hbar}{2e}I_B\hat{\phi}_{\rm R},\label{eq1}
\eea
where $H_{\rm 0}$ is the Hamiltonian of the PSC plus the kinetic inductance and the coupling with the current bias. Here $I_{\rm c}$ is the critical current through the PSC. We model the dissipation arising from the quasiparticle channel in the PSC and the external shunt resistor by incorporating two bosonic baths of harmonic oscillators. The bath related to the junction is $H_{\rm BP}$ and that with the shunt resistor is $H_{\rm BS}$.
\bea
H_{\rm BP}&=&\sum_{j=1}^{\infty}\Big[\f{\hat{P}_j^2}{2}+\f{\om_j^2}{2}\Big(\hat{X}_j-\f{c_j}{\om_j^2}\hat{\phi}\Big)^2\Big],\nn\\
H_{\rm BS}&=&\sum_{j=1}^{\infty}\Big[\f{\hat{p}_j^2}{2}+\f{1}{2}\tilde{\om}_j^2\hat{x}_j^2\Big],~\hat{\phi}_{\rm R}=\sum_{j=1}^{\infty}\lambda_j\hat{x}_j.
\eea
The operators $\hat{X}_j$, $\hat{x}_j$ and $\hat{P}_j$, $\hat{p}_j$ are conjugate positions and momenta of the bath oscillators with frequencies $\om_j$, $\tilde{\om}_j$. The coupling between the oscillators position and the phase operators are chosen to have an adequate description of noises from the baths. We choose ohmic dissipation from the baths by fixing the coupling parameters $c_j$, $\lambda_j$ and frequencies $\om_j$, $\tilde{\om}_j$ \cite{Guichard10}. Following the procedure in Ref. \onlinecite{Guichard10}, we employ the Heisenberg equation to write equations of motion for $\hat{\phi}$, $\hat{Q}$, $\hat{X}_j$, $\hat{x}_j$, $\hat{P}_j$ and $\hat{p}_j$. We integrate out the baths' degrees of freedom from these equations by the exact solutions of the baths' degrees of freedom along with the ohmic responses from the baths. Thus we obtain the following quantum Langevin equations for $\hat{\phi}$ and $\hat{I}_1$,
\bea
&&\hat{I}_1=I_{\rm c}\sin \hat{\phi} +\frac{\hbar }{2e R_{\rm PSC}}\f{d\hat{\phi}}{dt}+\frac{\hbar }{2e}C_{\rm PSC}\f{d^2\hat{\phi}}{dt^2}+\hat{I}_{\text{nP}},\label{eq4}\\
&&L_{\rm K}\frac{d\hat{I_1}}{dt}+\frac{\hbar }{2e }\frac{d\hat{\phi}}{dt}=\left(I_{\rm B}-\hat{I}_1\right)R_{\rm s}+ \hat{V}_{\rm ns}. \label{eq5}
\eea
Here, $\hat{I}_{\rm nP}$ and $\hat{V}_{\rm ns}$ are Gaussian current and voltage  noises from the PSC and the external resistor. The noise properties are, $\la \hat{I}_{\rm nP}\ra=0$, $\la \hat{V}_{\rm ns}\ra=0$ and the spectral function
\bea
S_{nP}(\om)&=&\int dt \f{e^{i\om t}}{2}\la \{ \hat{I}_{\rm nP}(t),\ \hat{I}_{\rm nP}(0)\}\ra=\f{\hbar \om}{R_{\rm PSC}}{\rm coth}\Big(\f{\hbar \om}{2k_{\rm B}T}\Big),\nn\\S_{ns}(\om)&=&\int dt \f{e^{i\om t}}{2}\la \{ \hat{V}_{\rm ns}(t),\ \hat{V}_{\rm ns}(0)\}\ra=\hbar \om R_{\rm s}{\rm coth}\Big(\f{\hbar \om}{2k_{\rm B}T}\Big),\nn
\eea
where $\{,\}$ denotes the anticommutator. Here $k_{\rm B}$ is the Boltzmann constant and $T$ is the temperature of the wire. The self-heating in the nanowire for a small external shunt resistor is weak, and $T$ is equal to the electrode temperature. Though the solution of the above quantum Langevin equations  is important and necessary for studying the macroscopic quantum tunneling, here we are interested in the classical dynamics of the system. Thus we replace the operators by their expectation values, for example, $\hat{\phi}$ by $\phi$. We also choose high temperature limit of the noise correlations, and replace them by $S_{nP}(\om)=2k_{\rm B}T/R_{\rm PSC}$ and $S_{ns}(\om)=2k_{\rm B}TR_{\rm s}$. The classical noise correlations in the time domain are given by
\begin{eqnarray}
\label{eq6}\left\langle I_{\text{nP}}(t_1)I_{\text{nP}}(t_2)\right\rangle =2k_{\rm B}T R_{\rm PSC}^{-1}\delta (t_1-t_2), \\
\label{eq7}\left\langle I_{\text{ns}}(t_1)I_{\text{ns}}(t_2)\right\rangle =2k_{\rm B}T R_{\rm s}^{-1}\delta (t_1-t_2),
\end{eqnarray}
where we define $V_{\rm ns}=R_{\rm s}I_{\rm ns}$. Now we introduce the following dimensionless variables: $i_{\rm B}=I_{\rm B}/I_{\rm c}$, $i_1=I_1/I_{\rm c}$, $i_{\rm nP}=I_{\rm nP}/I_{\rm c}$, $i_{\rm ns}=I_{\rm ns}/I_{\rm c}$ are the normalized currents, $r_{\rm s}=R_{\rm s}/R_{\rm PSC}$ is the normalized external shunt resistance, $t'=\om_{\rm p}t$ is the normalized time with $\omega _{\rm p}=\sqrt{2e I_{\rm c}/(\hbar  C_{\rm PSC})}$, $l_{\rm K}=L_{\rm K}\om_{\rm p}/R_{\rm PSC}$ is the normalized inductance, and $\bar{T}=k_{\rm B}\om_{\rm p}T/(\beta R_{\rm PSC}I^2_c)$ is the normalized temperature. 
Next we rewrite the classical version of the equations of motion Eqs. (\ref{eq4}) and (\ref{eq5}) and the classical noise correlations in Eqs.(\ref{eq6}) and (\ref{eq7}) using the dimensionless variables.
\begin{eqnarray}
\label{eq8} i_1&=&\sin  \phi +\beta \f{d\phi}{dt'}+\f{d^2\phi }{dt'^2}+i_{\rm nP},\\
\label{eq9} l_K\f{di_1}{dt'}+\beta \f{d\phi }{dt'}&=&\left(i_{\rm B}-i_1\right)r_{\rm s}+ i_{\text{ns}} r_s,\\
\left\langle i_{\text{nP}}(t_1')i_{\text{nP}}(t_2')\right\rangle &=&2 \beta  \bar{T} \delta (t_1'-t_2'),\\
\left\langle i_{\text{ns}}(t_1')i_{\text{ns}}(t_2')\right\rangle &=& 2 \beta r_s^{-1}\bar{T}\delta (t_1'-t_2'),
\end{eqnarray}
where $\beta=\sqrt{\hbar/(2eI_{\rm c}R^2_{\rm PSC}C_{\rm PSC})}$ is the damping factor. For typical parameters of MoSi superconducting nanowires\cite{Brenner12}, we have $R_{\rm PSC}\approx 100\ \rm{\Omega}$, $C_{\rm PSC}=8\ \rm{pF}$, $I_{\rm c}=5.55\ \rm{\mu A}$ and $\omega_{\rm p}=50\ \rm{GHz}$, $\beta=0.1$. As $\beta\ll 1$, thus the dynamics of the PSC is underdamped. For $l=90\ \rm{nm}$, $A=100\ \rm{nm^2}$, $\lambda=700\ \rm{nm}$ \cite{Plourde02}, we estimate the inductance $l_{\rm K}\approx 1$. Therefore, the kinetic inductance is not negligible in the practical nanowires.
We can find $i_1$ from Eq.(\ref{eq9}),
\bea
i_1(t_1') &=& i_1(t_1'=0)e^{-t_1'/\tau} \nn\\
 &+&\int _0^{t_1'}\f{e^{(t_2'-t_1')/\tau }}{\tau}\big(i_{\rm B}+i_{\text{ns}}-\f{\beta}{r_{\rm s}} \f{d\phi }{dt_2'}\big)dt_2',\label{eq12}
\eea
with $\tau=l_{\rm K}/r_{\rm s}$. After substituting Eq. (\ref{eq12}) in Eq. (\ref{eq8}) and using $i_1(t_1'=0)=i_{\rm B}$, we obtain a generalized quantum Langevin equation,
\begin{equation}\label{eq13}
\f{d^2\phi }{{dt_1'}^2}+\int _0^{t'_1}\gamma (t_1'-t_2') \f{d\phi }{dt_2'}dt_2'+\sin  \phi -i_{\rm B}+\tilde{i}_{\rm n}=0,
\end{equation}
where
\bea
\gamma (t_1'-t_2') &=&\beta \left[\delta (t_1'-t_2')+\frac{e^{-(t_1'-t_2')/\tau }}{\tau  r_{\rm s}}\right],\nn\\
\tilde{i}_{\rm n}&=&i_{\text{nP}}-\int _0^{t_1'}\frac{e^{(t_2'-t_1')/\tau }}{\tau}i_{\text{ns}}(t_2')dt_2'.
\eea
The presence of the inductor introduces a memory effect in the damping coefficient even though we choose ohmic dissipation from the baths at finite temperature. The fluctuation-dissipation theorem is still valid for the nonlocal dissipation and the noise, $\left\langle \tilde{i}_{\rm n}\left(t_1'\right)\tilde{i}_{\rm n}\left(t_2'\right)\right\rangle = 2 \bar{T} \gamma(t_1'-t_2')$.

\section{$V$-$I$ characteristics at $T=0$}
Here we investigate the effect of the kinetic inductance on the $V$-$I$ characteristics  in the absence of noises. Equation (\ref{eq8}) describes the dynamics of a phase particle in a tilted washboard potential, $U(\phi)=1-\cos\phi-i_1\phi$. When $i_{\rm B}$ is ramped up adiabatically, the voltage induced by this slowly increasing current is negligible, and thus $i_1=i_{\rm B}$. The phase particle always stays at the energy minimum of the potential at $\phi=\sin^{-1}i_{\rm B}$ until such a minimum disappears at $i_{\rm B}\ge 1$. Then the system switches to the resistive state. Therefore, the switching current $I_{\rm sw}$ is equal to $I_{\rm c}$ for a slowly varying bias current, and $I_{\rm sw}$ is independent of $l_{\rm K}$. The inductance $l_{\rm K}$ is only effective in the dynamic state, and the switching to the resistive state is unaffected by the presence of the kinetic inductor at $T=0$.

\begin{figure}[t]
\includegraphics[width=\columnwidth]{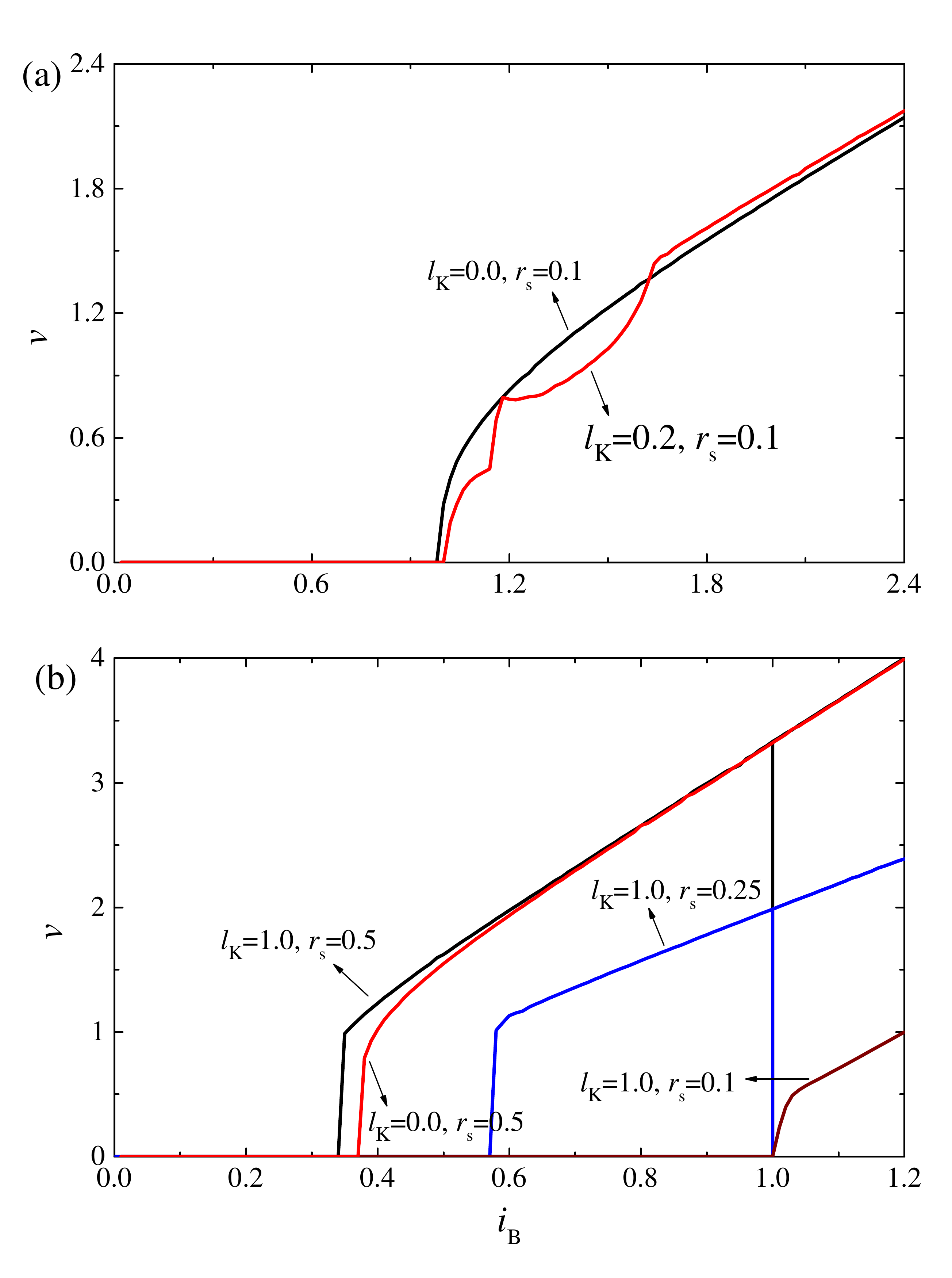}
\caption{(color online) $V$-$I$ characteristics at $T=0$ obtained by numerically solving Eqs. (\ref{eq8}) and (\ref{eq9}) without the noises for different $l_{\rm K}$ and $r_s$. We use $\beta=0.1$. Panel (a) shows the appearance of the kinks in the $V$-$I$ curve of an overdamped nanowire in the presence of a finite kinetic inductance.}
\label{f2}
\end{figure}

The $V$-$I$ curve becomes hysteretic for underdamped JJs, and the junction is retrapped to the superconducting state at $I_{\rm r}$, where the resistive state becomes unstable. The presence of the inductor changes the dynamics of the superconducting phase. For a small inductance $\tau=l_{\rm K}/r_{\rm s}\ll 1$, we obtain from Eq. (\ref{eq9}) to a leading order in $\tau$,
\[
i_1=i_{\rm B}-\f{\beta }{r_{\rm s}}\f{d\phi}{dt'}+\f{\beta \tau }{r_{\rm s}}\f{d^2\phi }{{dt'}^2}.\] 
Substituting $i_1$ into Eq. (\ref{eq8}), and rescaling the time $t'= t''\sqrt{\left(1-{\beta }\tau/{r_{\rm s}} \right)} $, we arrive at 
\begin{equation}\label{eq14}
i_B=\sin  \phi +\frac{\left(\beta +{\beta }/{r_{\rm s}}\right) }{\sqrt{1-{\beta }\tau/r_{\rm s} }}\f{d\phi}{dt''}+\f{d^2\phi }{{dt''}^2}.
\end{equation}
The retrapping current increases with the damping coefficient in the RCSJ model. The presence of the inductor increases the damping, thus the retrapping current is increased with an increasing inductance for a small $l_{\rm K}\ll 1$.

\begin{figure}[t]
\includegraphics[width=\columnwidth]{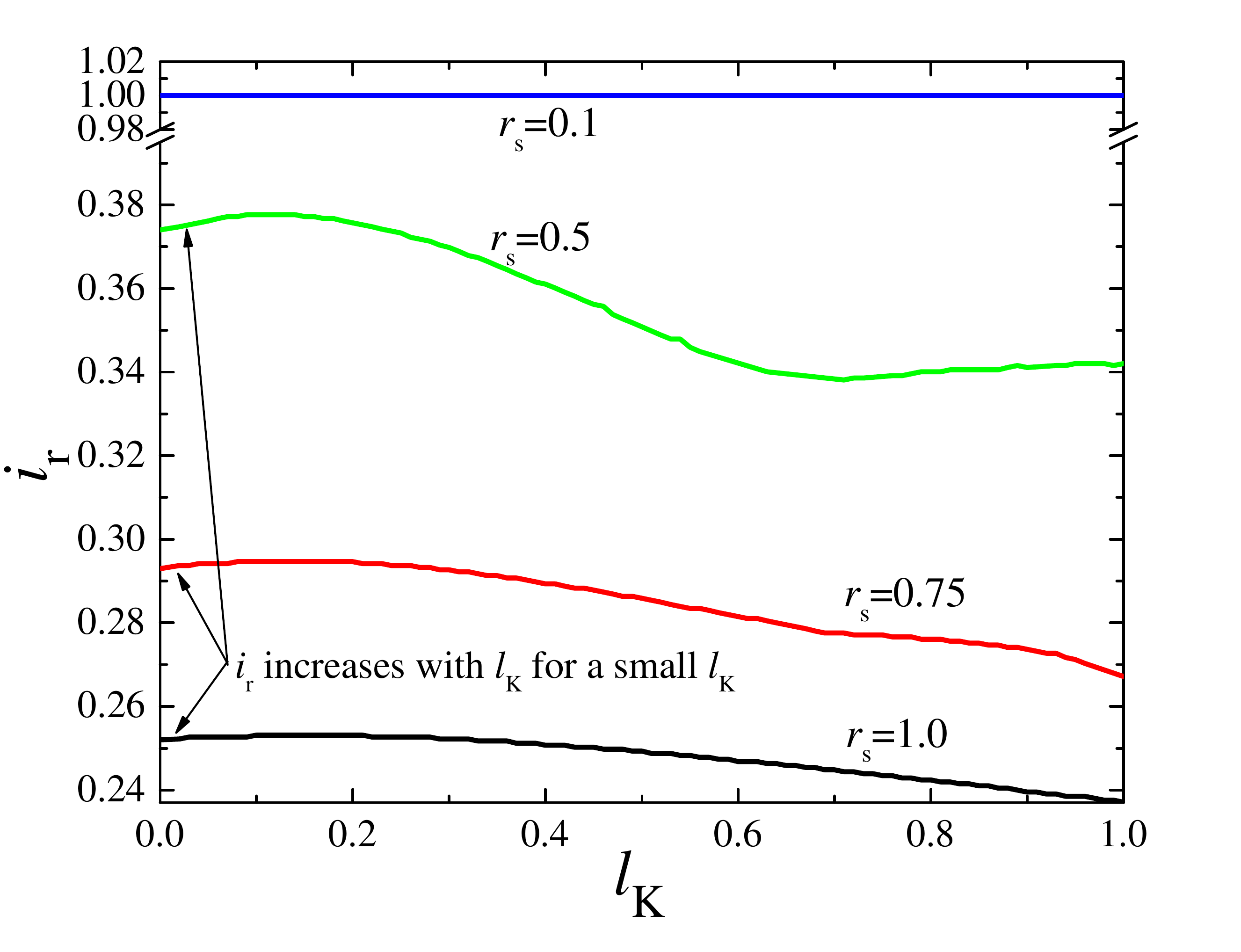}
\caption{(color online) Dependence of the zero-temperature retrapping current $i_{\rm r}$ on inductance $l_{\rm K}$ for different shunt resistances $r_{\rm s}$. Here $\beta=0.1$. }
\label{f3}
\end{figure}

We solve Eqs. (\ref{eq8}) and (\ref{eq9}) numerically for an arbitrary value of $l_{\rm K}$. The scaled voltage $v=V/(R_{\rm PSC}I_{\rm c})$ is calculated from $d\phi/dt'$. The typical $V$-$I$ curves are shown in Fig. \ref{f2} in the absence and presence of an inductance for different shunt resistances. In the overdamped region $r_s\ll 1$, some kinks develop in the $V$-$I$ curve when the kinetic inductance of SCNW is taken into account, as shown in Fig. \ref{f2} (a). These kinks were observed in experiments with shunted nanowires \cite{Brenner12} and it was argued there to come from the inductance of the resistive circuit. Here we show that a finite kinetic inductance of the SCNW can also gives rise such kinks. The dynamics of nanowire becomes underdamped for $\beta<1$ as $r_s$ increases, and a hysteresis in the $V$-$I$ curve shows up. The zero-temperature retrapping current $i_r$ depends on the inductance while the zero-temperature switching current $i_{sw}$ is independent of $l_K$. At a given $l_K$, $i_r$ increases with a decreasing $r_s$ and the hysteresis finally vanishes for a small enough $r_s$. The dependence of the zero-temperature $i_{\rm r}$ on $l_{\rm K}$ is shown in Fig. \ref{f3}. For a small $r_s$, $i_r=i_{sw}=1$ and the retrapping current is independent of $l_K$. For a large $r_s$, the retrapping current first increases with an increasing $l_{\rm K}$ for a small $l_{\rm K}$ which is consistent with the analysis for a small $l_{\rm K}$ in Eq. (\ref{eq14}). For a comparatively larger $l_{\rm K}$, the zero-temperature $I_{\rm r}$ decreases with an increasing $l_{\rm K}$.  The presence of an inductor does not affect the zero-temperature $I_{\rm sw}$, and the resistance of the resistive nanowire. The dynamical resistance is $R_{\rm d}\equiv dV/dI_{\rm B}\approx R_{\rm PSC}[\beta(1+1/r_{\rm s})]^{-1}$ in the resistive state. 

\begin{figure}[t]
\includegraphics[width=\columnwidth]{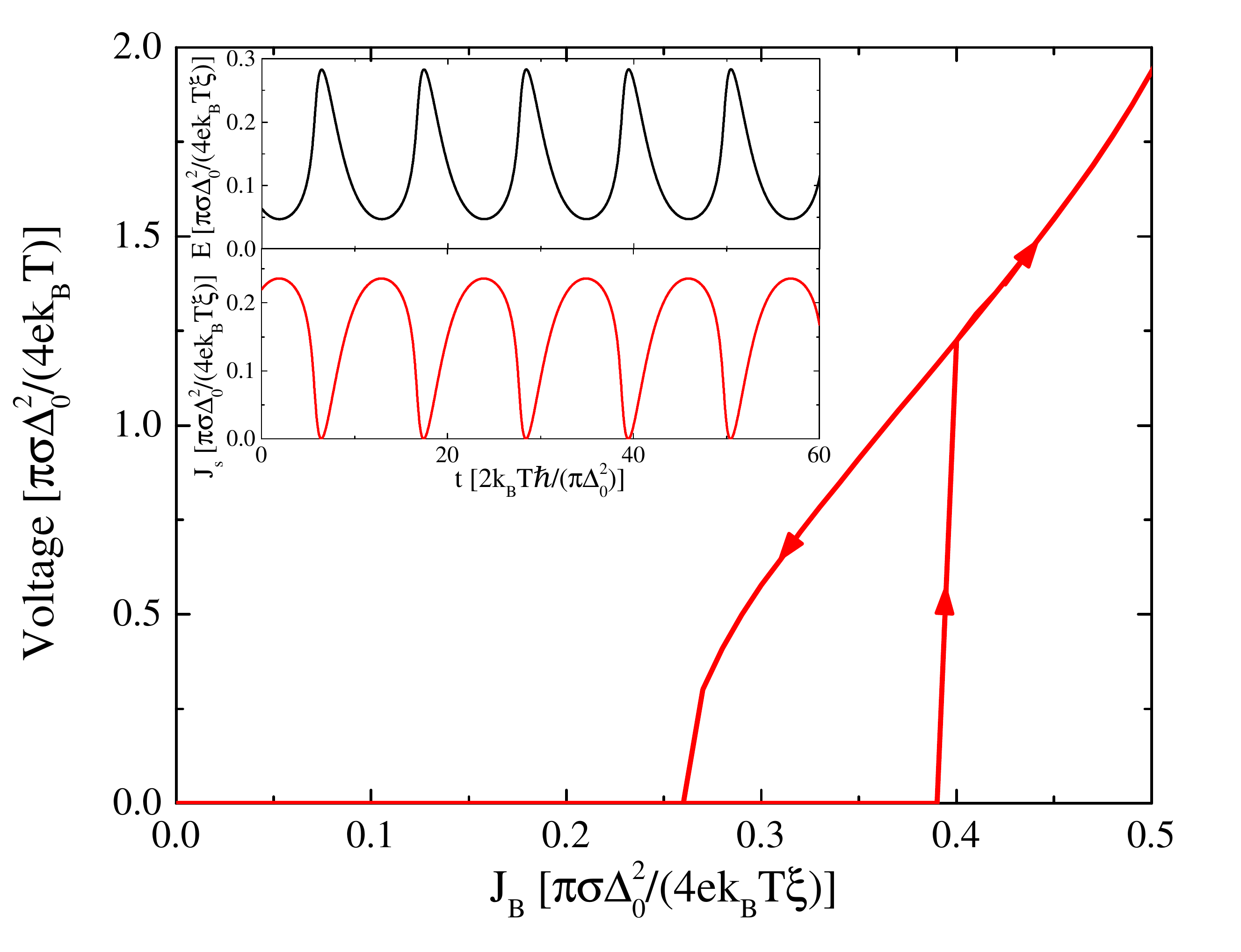}
\caption{(color online) \emph{V-I} curve for a SCNW biased by a dc curent $J_B$ obtained by numerical calculations of Eqs. \eqref{eqTDGL1} and \eqref{eqTDGL2}. Insets are for the time-dependent electric field and supercurrent across the PSC in the voltage state at $J_B=0.3$. Here $\Gamma=0.1$ and the length of the SCNW is $L=10$.}
\label{f0}
\end{figure}

Now some justifications and clarifying remarks on our use of the zero-dimensional RCSJ model to study one dimensional nanowire, seem to be in order. It has been demonstrated for a dc current-biased SCNW  that the uniform superconducting state becomes unstable at certain current values, and a phase slip center is developed. The dynamics of the superconducting order parameter $\Psi=\Delta\exp(i\theta)$ is governed by the time-dependent Ginzburg-Landau (TDGL) equations \cite{Ivlev84},
\begin{align}\label{eqTDGL1}
 \nonumber- u{\left(\frac{{|{\Psi ^2}|}}{{{\Gamma ^2}}} + 1\right)^{ - 1/2}}\left({\partial _t}\Psi  + {{i}}\varphi \Psi  + \frac{1}{2}{\Gamma ^{ - 2}}\Psi {\partial _t}|{\Psi}|^2\right)\\
  + \partial _x^2\Psi  + \Psi  - |{\Psi}|^2\Psi  = 0,
\end{align}
\begin{align}\label{eqTDGL2}
J_B =  - {\partial _x}\varphi  + \frac{1}{{2i}}({\Psi ^*}{\partial _x}\Psi  - \Psi {\partial _x}{\Psi ^*}),
\end{align}
where $u=5.79$, $\Gamma$ characterizes the pair breaking effect (for example, due to inelastic electron-phonon scattering), $J_B$ is the bias current and $\varphi$ is the electric potential. The first term at the right-hand side of Eq. \eqref{eqTDGL2} is the quasiparticle current, and the second term is the supercurrent. We use dimensionless units in writing down Eqs. \eqref{eqTDGL1} and \eqref{eqTDGL2}. \cite{Ivlev84} The length is in unit of $\xi$, the time is in unit of $2 k_B T\hbar/(\pi\Delta_0^2)$ and the current is in unit of $\pi\sigma\Delta_0^2/(4e k_B T \xi)$. Here $\xi$ is the coherence length, $\Delta_0$ is the energy gap in equilibrium and $\sigma$ is the quasiparticle conductivity. We solve Eqs. \eqref{eqTDGL1} and \eqref{eqTDGL2} numerically to obtain the \emph{V-I} curve and the time-dependent electric field and supercurrent across the PSC which are shown in Fig. \ref{f0}. The \emph{V-I} curves resemble that of a underdamped JJ. \cite{Hu10} Both the supercurrent and electric field oscillate in the voltage state. To the first harmonic approximation, the supercurrent in the SCNW $J_s$ can be written as $J_s=\bar{J}_s+J_c\sin(\phi_{\rm{PSC}})$ with the dc supercurrent $\bar{J}_s\approx J_c/2$ and the amplitude of the oscillating supercurrent $J_c$. \cite{Tinkham96} The electric field is $E=\partial_t\phi_{\rm{PSC}}$ according to the ac Josephson relation. Here $\phi_{\rm{PSC}}$ is the phase difference across the PSC. Then Eq. \eqref{eqTDGL2} can be simplified into as 
\begin{equation}\label{eqTDGL3}
\partial_t^2\phi_{\rm{PSC}}+\beta \partial_t\phi_{\rm{PSC}}+J_c\sin(\phi_{\rm{PSC}})=J_B-\bar{J}_s.
\end{equation}
The relaxation of the amplitude $\Delta$ of the order parameter $\Psi$ by pair breaking mechanism included in Eq. \ref{eqTDGL1} introduces an intrinsic inertia in the system, and it is phenomenologically incorporated by the $\partial_t^2\phi_{\rm{PSC}}$ term in Eq. \ref{eqTDGL3}. If one assumes that the amplitude is fixed during the time evolution, i.e., $\Delta(t)=\Delta_0$, then the term $\partial_t^2\phi_{\rm{PSC}}$ would be absent, and the resulting equation of motion would not show a hysteric behaviour in the \emph{V-I} curve as obtained in Fig. \ref{f0}. The above discussion does not yield an explicit expression for $\beta$ in terms of the physical quantities in Eqs. \eqref{eqTDGL1} and \eqref{eqTDGL2}. Nevertheless, it can be obtained by fitting the \emph{V-I} curves derived from Eq. \eqref{eqTDGL3} with the experiments. \cite{Brenner12} In principle one needs to employ the TDGL description in Eqs. \eqref{eqTDGL1} and \eqref{eqTDGL2} to discuss the dynamics of the superconductivity in SCNW. However we here show that the voltage-current characteristics obtained from the TDGL description is quite similar to that obtained from phenomenological description using the RCSJ model. Further the successful fitting of the experimental data based on the RCSJ model in Eq. \eqref{eqTDGL3} in Ref. \onlinecite{Brenner12} is encouraging. Therefore, we have adopted the JJ description in our present study. Please note that the effective bias current in the JJ model is $J_B-\bar{J}_s$. 
\section{Stochastic switching at $T > 0$}
At a finite temperature, upon ramping up the current, the system may switch to the resistive state at a bias current below the mean-field switching current $I_{\rm sw}=I_{\rm c}$ due to thermal fluctuations.  The distribution of $I_{\rm sw}$ is determined by  thermally activated escape rate of the phase particle from the meta-stable potential. The presence of an inductor induces a memory effect into the dynamics. In this section, we study the effect of the inductance on the rate, and then calculate the distribution of the switching current from the rate, which enables a direct comparison between the theory and the experiment.

\begin{figure}[t]
\psfig{figure=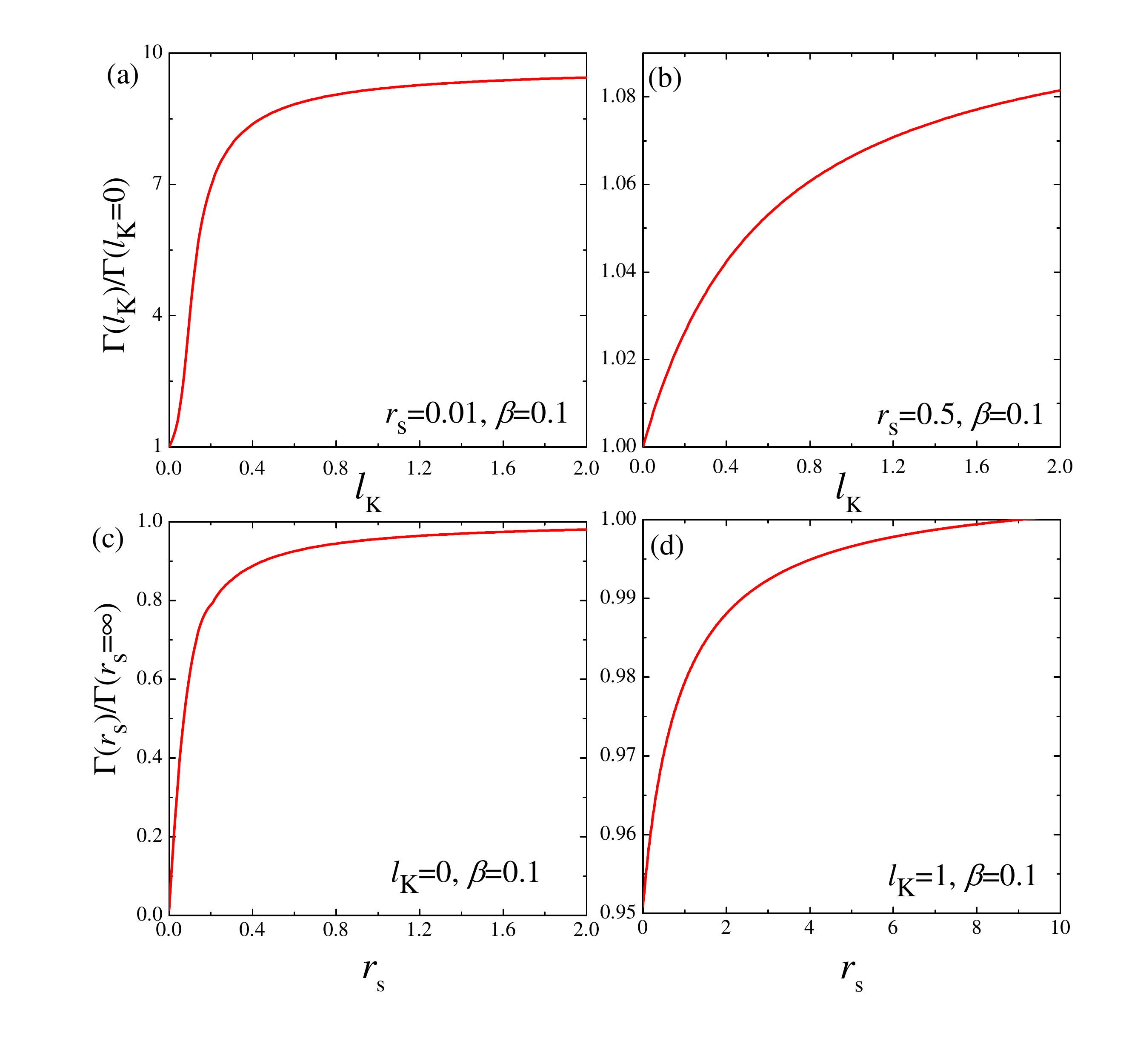,width=\columnwidth}
\caption{\label{f4}(color online) (a) and (b): Normalized switching rate $\Gamma(l_K)/\Gamma(l_K=0)$ as a function of inductance $l_K$ for two different shunt resistance $r_s$. (c) and (d): Normalized switching rate $\Gamma(r_s)/\Gamma(r_s=\infty)$ as a function of $r_s$ for two different $l_K$. The normalized switching rates are temperature independent. }
\end{figure} 

\begin{figure}[b]
\psfig{figure=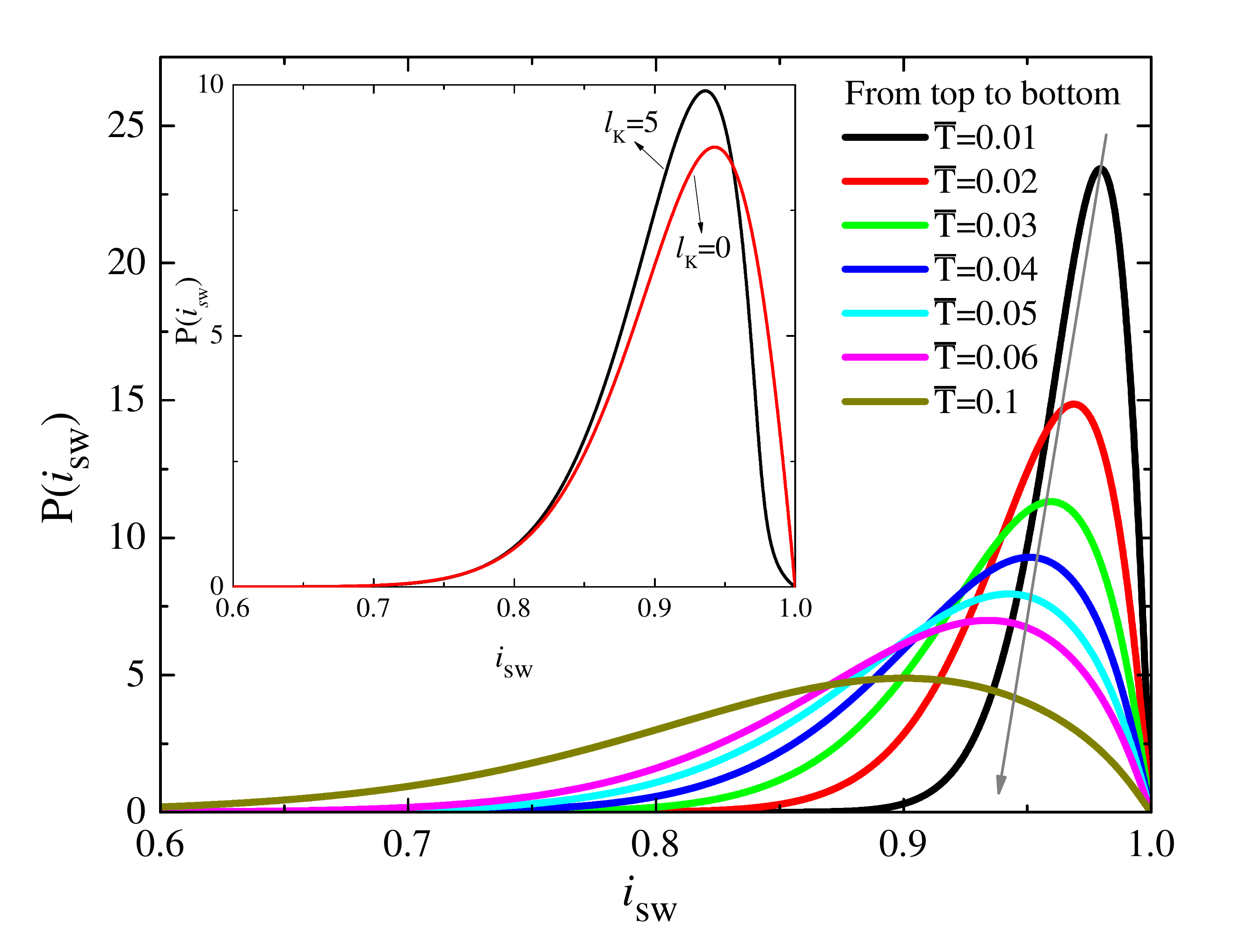,width=\columnwidth}
\caption{(color online) Distribution of the switching current $i_{sw}$ at different temperatures $\bar{T}$.  The inset shows the distributions of $i_{sw}$ with and without the inductor. The parameters for the main panel are $r_s=0.5$, $\beta=0.1$, $l_K=4$ and $\dot{i}_B=0.01$, and for the inset is $r_s=0.02$, $\beta=0.2$, $\bar{T}=0.04$ and $\dot{i}_B=0.01$. } \label{f5} 
\end{figure}

When $l_{\rm K}$ is small, $\tau=l_{\rm K}/r_{\rm s}\ll 1$, Eq. (\ref{eq13}) up to a linear order in $\tau$ becomes
\begin{equation}\label{eq15}
i_{\rm B}=\sin  \phi +\left(\beta +\frac{\beta }{r_{\rm s}}\right) \f{d\phi }{dt'}+\left(1-\frac{\beta }{r_{\rm s}}\tau \right)\f{d^2\phi }{{dt'}^2}+\tilde{i}_{\rm n}, 
\end{equation}
\[
\left\langle\tilde{i}_{\rm n}\left(t'_1\right)\tilde{i}_{\rm n}\left(t'_2\right)\right\rangle = 2 \bar{T} \left(\beta +\frac{\beta }{r_{\rm s}}\right)\delta(t'_1-t'_2).
\]
Equation (\ref{eq15}) describes the motion of  a particle with mass $M=(1-\beta\tau/r_{\rm s})$ in the potential $U(\phi)$, and the thermal escape rate $\Gamma$ for this problem is well known\cite{Hanggi90},
\begin{equation}\label{eq16}
\Gamma=\frac{\sqrt{{\gamma_1 ^2}+4M\omega _b^2}-\gamma_1 }{2M\omega _b}\frac{\omega _0}{2\pi }\exp [-\Delta U/\bar{T}],
\end{equation}
where $\gamma_1=(\beta+\beta/r_{\rm s})$,  $\Delta U=U(\phi_b)-U(\phi_0)$, $\omega_b=\sqrt{-\partial_{\phi}^2 U(\phi_b)}$, and $\omega_0=\sqrt{\partial_{\phi}^2 U(\phi_0)}$. Here $\phi_b$ ($\phi_0$) is the phase at the barrier (well). 

For a general value of $\tau$, the switching rate with a nonlocal damping coefficient has been calculated in Ref. \onlinecite{Hanggi82}, and we here use their results,
\begin{equation}\label{eq17}
\Gamma=\frac{\omega _0\alpha }{2\pi  \omega _b}\exp [-\Delta U/\bar{T}],
\end{equation}
where $\alpha$ is a pole at the right-half plane of the function $f(z)=[{z^2-\omega _b^2+z {\rm L}[\gamma (t')]}]^{-1}$, where ${\rm L}[\gamma (t')]=\beta +{\beta }/[({1+z \tau })({r_{\rm s}})]^{-1}$ is Laplace transform of $\gamma(t')$.  The kinetic inductance does not change the barrier height $\Delta U$, but modifies the pre-exponential factor of the rate. The rate is still dominated by the barrier, but the enhancement of the rate can be large by taking the inductor into account in some parameter space. For a small $\tau$, Eq.(\ref{eq17}) reduces to Eq. (\ref{eq16}), while for $\tau\rightarrow \infty$
\begin{equation}\label{eq18}
\Gamma=\frac{\sqrt{{\beta ^2}+4\omega _b^2}-{\beta }}{2\omega _b}\frac{\omega _0}{2\pi }\exp [-\Delta U/\bar{T}].
\end{equation}
In the last limit, the effect of kinetic inductor and shunt resistor vanishes and the rate is the same as that in a single JJ. The dependence of the switching rate on $l_{\rm K}$ at two different $r_{\rm s}$ is shown in Fig. \ref{f4} (a) and (b). The rate increases with $l_{\rm K}$. For $l_{\rm K}=0$, the effect of the shunt resistor is to renormalize the damping of the nanowire $\beta_{\rm{eff}}=\beta+\beta/r_{\rm s}$, which reduces the rate. When $l_{\rm K}$ is quite large, the current in the wire does not change much with time, $i_1 \sim i_{\rm B}$ and the shunt circuit becomes less important. Thus the switching rate reaches the limit of a JJ without the shunt resistance. For $R_{\rm s}/R_{\rm PSC}\sim 1$, the increment of the rate is small, while for a small shunt resistance $R_{\rm s}/R_{\rm PSC}\ll 1$, the enhancement of the switching rate can be large due to the presence of the inductor. The dependence of the switching rate on $r_{s}$ at two different $l_{K}$ is shown in Fig. \ref{f4} (c) and (d). The rate increases with $r_s$, but the increment is small for a large $l_K$.

We proceed to calculate the distribution of the switching current from the rate. In  experiments, one sweeps the bias current slowly to avoid the nonequilibrium heating effect (much slower than any other characteristic time scale in the system), and measures the voltage. Once there is a nonzero voltage signal, a phase slip event is counted. One then resets the system to the superconducting state by reducing the current, and repeats the measurements, from which the distribution of switching current is constructed.  The probability for the system leaving the meta-stable potential after time $t$ during the process of ramping up the bias current is\cite{Fulton74}
\begin{equation}\label{eq19}
W\left[i_B(t')\right]=1-\exp \left[-\int_0^{t'} \Gamma \left[i_B(t_1')\right] dt_1'\right].
\end{equation}
When $i_B$ is increased at a constant rate $\dot{i}_B\equiv d{i}_B/dt'=\rm{constant}$, the distribution function is then given by
\begin{equation}\label{eq20}
P\left(i_B\right)=\frac{dW}{d i_B}=\frac{\Gamma \left[i_B\right]}{\dot{i}_B}\exp \left[-\int_0^{i_B} \frac{\Gamma \left[i_B'\right]}{\dot{i}_B}  di_B'\right].
\end{equation}
The distributions of $i_{\rm sw}$ for several temperatures are shown in Fig. \ref{f5}. The linewidth of the distribution increases with an increasing temperature.  The distributions in the presence and absence of an inductor are shown in the inset of Fig. \ref{f5}. The shape of the switching current distribution is modified in the presence of the inductor.

\section{Discussion}
Now we discuss the effect of self-heating on the $V$-$I$ characteristics. The heat generated in the unshunted nanowires is removed through the electrodes that are attached to the nanowires. The heating effect in the unshunted nanowires plays an important role in determining the transport properties of the nanowires, and one has to take into account the heat-diffusion process in order to understand the $V$-$I$ characteristics. For the shunted nanowires, the bias current redistributes in the shunt circuit once the phase slip occurs in the nanowires. This redistribution is fast and depends on the kinetic inductance of the nanowires and the shunt resistance, $\tau_1=L_K/(R_s+R_{\mathrm{PSC}})$ as discussed in Sec.II. The phase slip process is then suppressed exponentially because of the small bias current in the nanowire, which allows the nanowire to cool down. Once the nanowire  switches back to the superconducting state, the bias current flows again into the nanowire within a time scale $\tau_2=L_K/R_s$ that is much longer than $\tau_1$. For typical nanowires, $\tau_2 \approx 10$ ns. Thus, there is sufficient time for the nanowires to cool down if the phase slip rate is smaller than the rate of current redistribution, $1/\tau_2$. Therefore the self-heating effect for the shunted nanowires is minimized, as was also confirmed experimentally. \cite{Brenner12}

In short, we have proposed a new effective circuit for externally shunted superconducting nanowires  by taking the kinetic inductor of the superconductor into account. The presence of inductor introduces a nonlocal damping in the dynamics of superconductivity. (a) First it minimizes the heating in the nanowire by introducing a delay in the redistribution of current between the resistive shunt and the nanowire. (b) The kinetic inductor does not affect the mean-field switching current, but it modifies the mean-field retrapping current in a non-trivial way. (c) We also consider the thermally assisted switching from the superconducting state to the resistive state. The switching rate increases with the inductance. The resulting distribution of the switching current is also modified. 

Inductive shunting plays a significant role in building superconducting circuits using Josephson junctions \cite{Manucharyan09}. One of the interesting direction for potential application of SCNWs is to use them as superconducting qubits in quantum information processing and circuit quantum electrodynamics type set-ups. Therefore, the inherent kinetic inductance of SCNWs may provide an advantage for using the nanowires compared to the Josephson junctions.  A voltage biased narrow SCNW with coherent quantum phase slips is exact dual to a current biased RCSJ model \cite{Mooij06}. It has been argued recently that weak links with slightly larger resistivity occur naturally in apparently homogeneous nanowires, and they can localize quantum phase slips by quantum fluctuations at low temperatures \cite{Vanevic12}. Here we show that the phase slip center in series with an inductor is responsible for the dynamics in the externally shunted SCNWs at temperatures close to the critical. Since the inductance depends on the length and the cross section of the nanowire, the role of the kinetic inductor can be tested in experiments by measuring the voltage-current characteristics and the distributions of the switching current in resistively shunted nanowires of different sizes.  

\section{Acknowledgments}   

One of the authors (S.Z.L.) is grateful to L. N. Bulaevskii for helpful discussions.  S.Z.L. was supported by the U.S. Department of Energy, Office of Basic Energy Sciences, Division of Materials Sciences and Engineering. D.R. gratefully acknowledges the support of the U.S. Department of Energy through LANL/LDRD Program for this work.

\end{document}